\documentclass[aps,prl,floatfix,twocolumn,showpacs,superscriptaddress,groupedaddress]{revtex4-2}
\usepackage{graphicx}%
\usepackage{amsmath, amssymb}
\usepackage{color}
\usepackage{amsfonts}
\usepackage{hhline}
\usepackage{braket}
\usepackage{bm}
\usepackage{mathrsfs}
\usepackage{hyperref}
\hypersetup{colorlinks=true,breaklinks,linkcolor=blue,urlcolor=blue,citecolor=blue}
\def\vec#1{\boldsymbol #1}

\begin{document}
\title{
Compensated Ferrimagnets with Colossal Spin Splitting in Organic Compounds
}
\author{Taiki Kawamura$^1$}
\author{Kazuyoshi Yoshimi$^2$}
\author{Kenichiro Hashimoto$^3$}
\author{Akito Kobayashi$^1$} 
\author{Takahiro Misawa$^2$}
\email{tmisawa@issp.u-tokyo.ac.jp}
\affiliation{$^1$Department of Physics, Nagoya University, Nagoya, Aichi 464-8602, Japan \\
$^2$Institute for Solid State Physics, University of Tokyo, Kashiwa, Chiba 277-8581, Japan \\
$^3$Department of Advanced Materials Science, University of Tokyo, Kashiwa, Chiba 277-8561, Japan
}
\begin{abstract}
The study of the magnetic order has recently been invigorated by the discovery of exotic collinear antiferromagnets with time-reversal symmetry breaking. 
Examples include altermagnetism and compensated ferrimagnets, 
which show spin splittings of the electronic band structures even at zero net magnetization, 
leading to several {unique}  
transport phenomena, notably spin-current generation. Altermagnets demonstrate anisotropic spin splitting, such as $d$-wave, in momentum space, whereas
compensated ferrimagnets exhibit isotropic spin splitting. However, methods to realize compensated ferrimagnets are limited. Here, we demonstrate a method to realize a fully compensated ferrimagnet with isotropic spin splitting utilizing the dimer structures inherent in organic compounds. Moreover, based on $ab$ $initio$ calculations, we find that this ferrimagnet can be realized in the recently discovered organic compound (EDO-TTF-I)$_2$ClO$_4$. Our findings provide an unprecedented strategy for 
using the dimer degrees of freedom in organic compounds to realize 
fully compensated ferrimagnets with colossal spin splitting.
\end{abstract}

\maketitle

{\it Introduction.---}Collinear antiferromagnets have traditionally been viewed as conventional magnetic orderings that lack 
{unique} phenomena such as spin current generation. However, recent theoretical advances have identified exotic collinear antiferromagnets with time-reversal breaking, notably altermagnets~\cite{Smejkal_NatureReview20022,Smejkal_PRX2022,Smejkal_PRX2022b} and compensated ferrimagnets~\cite{Mazin_PRX2022}. These  
magnetic states exhibit spin splitting in their electronic band structures even without net magnetization. 
Because spin splitting may drive spin-dependent 
{novel} transport phenomena and unconventional superconducting phases, 
these 
{distinctive} antiferromagnets have attracted considerable attention.

Several materials have been proposed as candidates for altermagnetism~\cite{Noda_PCCS2016,Naka_NCom2019,Hayami_JPSJ2019,Ahn_PRB2019,Yuan_PRB2020,Smejkal_ScienceAd2020,Naka_PRB2020,Naka_PRB2021,Mazin_PNAS2021,Smejkal_PRX2022b}. For example, $\kappa$-ET-type organic compounds~\cite{Naka_NCom2019} and transition metal oxide RuO$_{2}$~\cite{Ahn_PRB2019,Smejkal_ScienceAd2020,Feng_NatureEle2022} exhibit altermagnetism with anisotropic spin splitting in electronic band structures.
These materials are expected to exhibit spin-dependent transport and anomalous Hall effects owing to anisotropic spin splitting. In contrast, compensated ferrimagnets offer isotropic spin splitting, increasing efficiency for spin-current generation. 
The concept of metallic compensated ferrimagnetism (half-metallic antiferromagnetism) was introduced by van Leuken and de Groot~\cite{Leuken_PRL1995}. Since then, several candidate materials have been suggested using $ab$ $initio$ calculations~\cite{Pickett_PRB1998,Akai_PRL2006}. 
More recently, monolayer MnF$_2$ was proposed as {an insulating} compensated ferrimagnet~\cite{Egorov_JPCS2021}. 
{Since the magnetic moment of compensated insulating ferrimagnets is strictly zero due to the Luttinger theorem~\cite{Mazin_PRX2022}, small perturbations do not change the compensation condition. In addition, compensated ferrimagnets have lower crystal symmetry than altermagnets~\cite{Mazin_PRX2022}. Therefore, compensated ferrimagnets have advantages over altermagnets, leading to various potential applications, such as thin-film synthesis.}
However, the number of compensated ferrimagnets discovered experimentally is limited~\cite{Stinshoff_AIP2017,Stinshoff_PRB2017,Midhunlal_JMMM2019,Semboshi_SciRep2022}.
To harness the compensated ferrimagnets, 
a simple method to realize them is necessary.

\begin{figure}[t]
\begin{center}
\includegraphics[width=90mm]{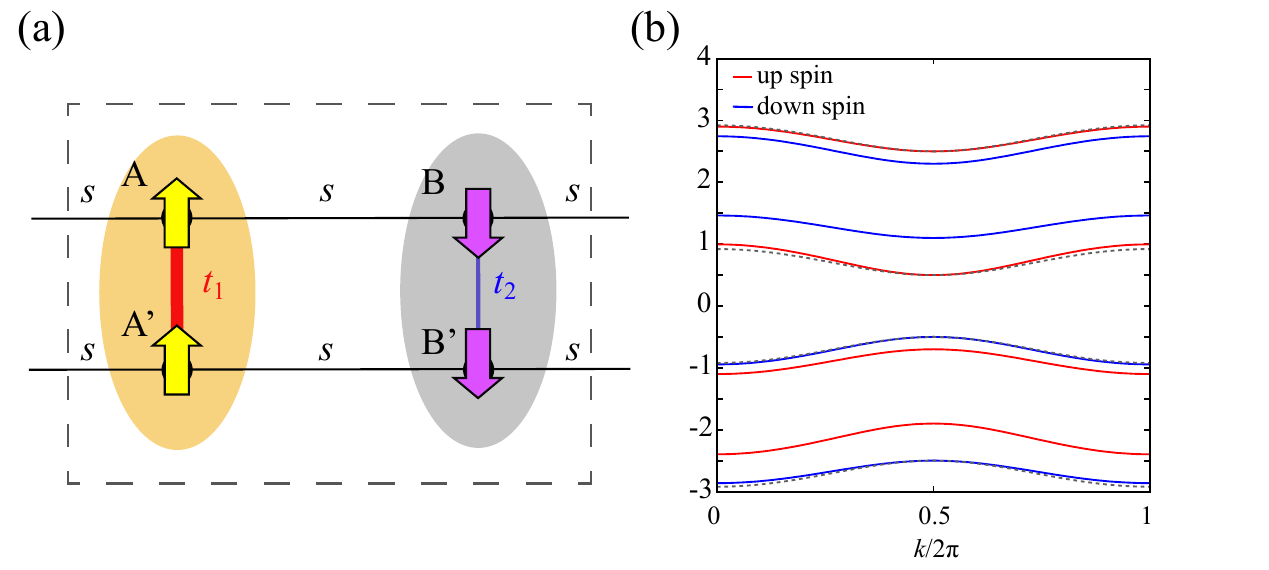}
\end{center}
\caption{ 
(a)~Schematic of a one-dimensional model showing the compensated ferrimagnetism.
The broken lines show a unit cell. 
{The inter-dimer hopping integral $s$, the intra-dimer hopping integrals $t_1$ and $t_2$ are represented by the horizontal thin black lines, 
the vertical thick red line, and the vertical thin blue line, respectively. The up (down) spin polarizations are described by the yellow upward (purple downward) arrows.} 
(b)~Band structures of the Hamiltonian defined in Eq.~(\ref{eq:Ham}).
We take $t_{1}=1.0$, $t_{2}=0.6$, $s=0.6$, $\delta=0.2$, and $\Delta=1.5$.
For comparison, we also show the band structures of the equivalent dimer case with
the broken curves, whose parameters are given by
 $t_{1}=t_{2}=1.0$, $s=0.6$, $\delta=0.0$, and $\Delta=1.5$.
}
\label{fig:lattice}
\end{figure}

In this Letter, we present a path towards fully compensated ferrimagnets with colossal spin splitting using typical dimer structures in organic compounds. Using a simple one-dimensional model, we demonstrate that a collinear antiferromagnetic order with inequivalent dimers can induce fully compensated ferrimagnets. Furthermore, we 
find that the recently discovered organic compound  (EDO-TTF-I)$_2$ClO$_4$
{[EDO-TTF-I=4,5-ethylenedioxy-4$^{\prime}$-iodotetrathiafulvalene]~~\cite{Y.Nakano.PC}}
can realize this mechanism based on $ab$ $initio$ calculations. The experiments demonstrate that
(EDO-TTF-I)$_2$ClO$_4$ undergoes a structural transition with anionic ordering at 190 K.
Below the structural transition, unit-cell doubling occurs with the extended unit cell containing two inequivalent dimers.
By deriving and solving the $ab$ $initio$ effective Hamiltonian for the low-temperature phase of (EDO-TTF-I)$_2$ClO$_4$, we find that the ground state is a collinear antiferromagnet with isotropic spin splitting, i.e., a fully compensated ferrimagnet. 

{\it Simple model.---}First, to show the key idea for realizing the compensated ferrimagnets, we consider a one-dimensional model whose unit cell contains
two {\it inequivalent} dimers. A schematic of 
the model is illustrated in Fig.~\ref{fig:lattice} (a).
The inequivalence between the two dimers is characterized using 
difference between the intra-dimer hoppings $t_{1}$ and $t_{2}$
and chemical potential difference $\delta$.
We also consider the collinear dimer antiferromagnetic (DAF) state, where the up-(down-)spin electrons are located on A and A$^{\prime}$ (B and B$^{\prime}$).
This DAF state is not invariant to any
combination of time reversal with translation/rotation
operations because of dimer inequivalence. 
Thus, isotropic spin splitting is expected in this dimer collinear DAF state.

To determine the mechanism of spin splitting in this model,
we consider the following tight-binding Hamiltonian for the DAF state: 
\begin{align}
&\mathcal{H}=\sum_{k,\sigma}\vec{c}^{\dagger}_{k\sigma}H_{\sigma}(k)\vec{c}_{k\sigma}, \\
&H_{\sigma}({k})=
\begin{pmatrix}
\sigma{\Delta}     ~~& t_1               ~~& A(k)~~& 0 \\
t_1                ~~& \sigma{\Delta}    ~~& 0                 ~~& A(k) \\
A(k)^{*}         ~~& 0                 ~~& -\sigma{\Delta}+\delta   ~~& t_2 \\
0                  ~~& A(k)^{*}    ~~& t_{2}             ~~& -\sigma{\Delta}+\delta
\end{pmatrix},
\label{eq:Ham}
\end{align}
where {$\vec{c}_{\vec{k}\sigma}^{\dagger}=(c_{A{k}\sigma}^{\dagger},c_{A^{\prime}{k}\sigma}^{\dagger},c_{B{k}\sigma}^{\dagger},c_{B^{\prime}{k}\sigma}^{\dagger})$},
$A(k)=s(1+e^{-ik})$
and $\Delta$ denotes the gap induced by the DAF order.
The spin index $\sigma$ takes $+1$ and $-1$ for the up- and down-spins, respectively.
The eigenvalues of this Hamiltonian are given by
\begin{align}
&E_{0,\sigma,\pm}(k)={\delta/2+}t_{+}\pm[(\sigma\Delta+t_{-}{-\delta/2})^2+2s^2 C(k)]^{1/2},\\
&E_{1,\sigma,\pm}(k)={\delta/2}-t_{+}\pm[(\sigma\Delta{-t_{-}-\delta/2})^2+2s^2 C(k)]^{1/2},
\end{align}
where $t_{\pm}=(t_{1}\pm t_{2})/2$ and
$C(k)=(1+\cos{k})$.
Thus, 
spin splitting of the bands is induced by $t_{-}=(t_{1}-t_{2})/2$ and $\delta$, i.e., inequivalence of the dimers.
From these expressions, it is evident that
the differences in intra-dimer hoppings 
have the 
{similar} gap-opening effect as the differences in the chemical potentials.
Because $t_{-}$ and $\delta$ are independent of the wave number, spin splitting is isotropic.

Figure~\ref{fig:lattice}(b) shows the electronic band structures of
the Hamiltonian defined in Eq.~(\ref{eq:Ham}) for a typical parameter set.
We also plot the band structures when the two dimers are equivalent 
($t_{1}=t_{2}$ and $\delta=0$).
As expected, isotropic spin splitting occurs in the band structures.
At commensurate filling (i.e., three-quarter, half, and quarter filling),
the DAF state is insulating, and the net magnetization is zero because the number of up- and down-bands below the Fermi energy are the same.
Thus, the DAF state at the commensurate filling is the 
fully compensated ferrimagnets with isotropic spin splitting.

\begin{figure}[tpb]
\begin{center}
\includegraphics[width=85mm]{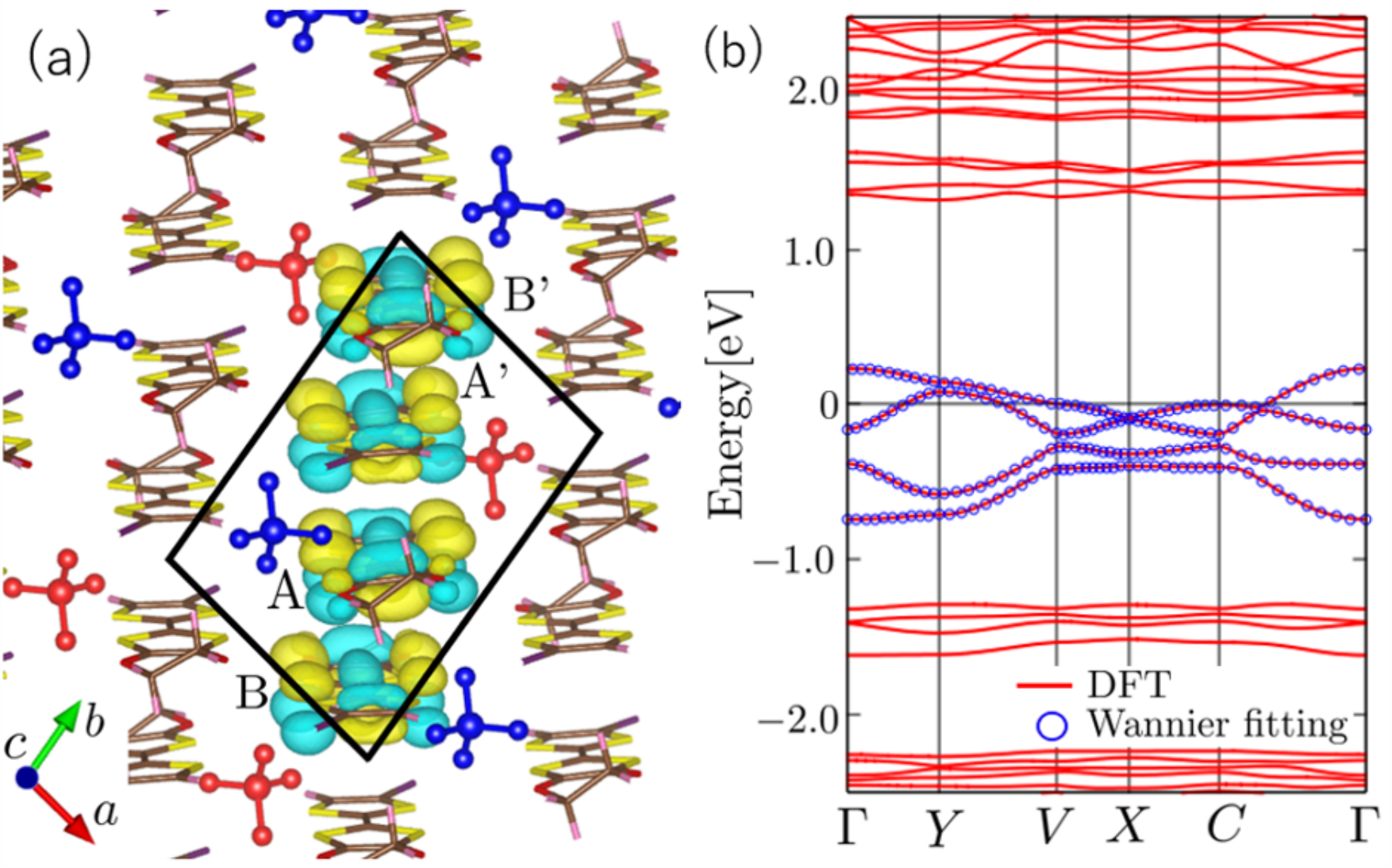}
\end{center}
\caption{(a) Crystal structure of (EDO-TTF-I)$_2$ClO$_4$ at $T$ = $100$ K and the real-space distribution of maximally localized Wannier functions (MLWFs) drawn by \texttt{VESTA}~\cite{VESTA}.
{Anions ClO$_{4}$ are represented by balls and rods. 
Colors of anions represent their orientations. Due to the different configurations of anions around the dimers, two dimers (A-A$^{\prime}$ and B-B$^{\prime}$) become inequivalent.}
Black lines show the unit cell including four molecules. 
(b) Energy band structure obtained by the DFT calculation (red lines) and the MLWFs (blue circles) {for the paramagnetic states}. The Fermi energy is set to zero.}
\label{Fig2}
\end{figure}

{\it Crystal and electronic structures of (EDO-TTF-I)$_{2}$ClO$_{4}$}.---
First, we summarize the crystal structure of (EDO-TTF-I)$_{2}$ClO$_{4}$.
(EDO-TTF-I)$_2$ClO$_4$ consists of EDO-TTF-I molecule with $+1/2$ charge ($3/4$ filling) and anion ClO$_4$ layers. 
Above $190$ K, the unit cell contains {two EDO-TTF-I molecules}
and space inversion symmetry is macroscopically protected because of the random orientation of ClO$_4$. 
By lowering the temperature, the structural phase transition 
with anion ordering occurs at approximately $T=190$ K,
which induces unit-cell doubling, as evidenced by the X-ray analysis.
As shown in Fig.~\ref{Fig2} (a), 
each unit cell contains four molecules in the low-temperature phase. 
A and A$^{\prime}$ (B and B$^{\prime}$) molecules in a unit cell form a dimer, 
 referred to as dimer I (dimer II). 
The inversion centers are at the centers of each dimer.
{The inequivalence of the dimers I and II can also be 
understood from the partial density of states (PDOS),
which are shown in the Supplemental Material~\cite{supplementary}.}
Because dimers I and II are inequivalent after the structural phase transition, (EDO-TTF-I)$_2$ClO$_4$ will exhibit compensated ferrimagnetism if 
appropriate antiferromagnetic order occurs.
We note the related organic compound (EDO-TTF-I)$_{2}$PF$_{6}$~\cite{Y.Nakano2018} has
similar crystal structures; however, all the dimers are equivalent.
This indicates that the ordering of ClO$_{4}$ plays an essential role in realizing 
inequivalent dimers. 

Next, we summarize the low-temperature electronic structure of  (EDO-TTF-I)$_2$ClO$_4$.
Although the resistivity exhibits semimetallic behavior immediately after the structural phase transition, a metal-insulator phase transition occurs at approximately $T=95$ K~\cite{Y.Nakano.PC}. 
It is confirmed that the metal-insulator transition does not accompany a structural change.
As explained later,
$ab$ $initio$ analysis suggests that 
the DAF ordering is the origin of the insulating phase
although the origin of the metal-insulator transition is not experimentally clarified yet.

{\it Ab initio effective Hamiltonian for (EDO-TTF-I)$_{2}$ClO$_{4}$.---}
To investigate the electronic structures of 
the low-temperature phase in (EDO-TTF-I)$_{2}$ClO$_{4}$,
we derive the $ab$ $initio$ low-energy effective Hamiltonian, 
based on band structures obtained by the density functional theory (DFT).
We use Quantum ESPRESSO~\cite{P.Giannozzi2017} to obtain the DFT bands.
In this study, we employ the optimized norm-conserving Vanderbilt pseudopotentials and plane-wave basis sets~\cite{Hamann_ONCV2013, Schlipf_CPC2015}. The exchange-correlation functional used in this study is the generalized gradient approximation proposed by Perdew, Burke, and Ernzerhof~\cite{GGA_PBE}.
The cut-off energy of the wave functions and charge densities are set to be $80$ Ry and $320$ Ry, respectively. 
During the self-consistent loop process, a $5\times 5\times 3$ uniform $\bm{k}$-point grid and {Methfessel-Paxton smearing method} are used~\cite{M-P}.
We use the crystal structure data at 100 K~\cite{Y.Nakano.PC}.
We perform structural optimization for hydrogen atoms 
and use the optimized structure in the following analyses.
{In the DFT calculations, we only consider the paramagnetic solutions.}
Figure \ref{Fig2} (b) shows the obtained energy band structures of (EDO-TTF-I)$_{2}$ClO$_{4}$. 
{The four bands around the Fermi level, which are isolated from the other bands, {mainly consist of the highest occupied molecular orbitals of}  
{the four EDO-TTF-I molecules in the unit cell (A,A$^\prime$,B, and B$^\prime$).}
We select {these four bands} as the low-energy degrees of freedom and use {them} to construct the maximally localized Wannier functions (MLWFs) using RESPACK~\cite{K.Nakamura2021}.}
Isosurfaces of the MLWFs are shown in Fig.\ref{Fig2} (a). 
In addition, we confirm that the bands interpolated by the MLWFs accurately reproduce the DFT band structures (Fig.\ref{Fig2} (b)).

After constructing MLWFs, we derive 
low-energy effective Hamiltonian, which is given by
\begin{align}
&H_{\rm EDO}=H_{0}+H_{\rm int},\notag \\
&H_{0}=\sum_{i,j,\alpha,\beta,\sigma}t_{i\alpha j\beta} c^{\dagger}_{i\alpha\sigma}c_{j\beta\sigma },\notag \\
&H_{\rm int}=\sum_{i,\alpha}U_{i\alpha}n_{i\alpha\uparrow}n_{i\alpha\downarrow} 
+\frac{1}{2}\sum_{i,j,\alpha,\beta}V_{i\alpha j\beta}N_{i\alpha}N_{j\beta},
\label{eq:lowham}
\end{align}
where $c^{\dagger}_{i\alpha\sigma}$ ($c_{i\alpha\sigma}$) is a creation (annihilation) operator for
an electron in the $i$-th unit cell with orbitals $\alpha$= A, A$^{\prime}$, B, B$^{\prime}$ 
and spin $\sigma$. The number operators are defined as 
$n_{i\alpha\sigma}= c^{\dagger}_{i\alpha\sigma}c_{i\alpha\sigma}$ and
$N_{i\alpha}=n_{i\alpha\uparrow}+n_{i\alpha\downarrow}$. 
The transfer integrals $t_{i\alpha j\beta}$ are evaluated using the MLWFs. 
We also evaluate the screened Coulomb interactions $U_{i\alpha}$ and $V_{i\alpha j\beta}$ 
using the constrained random phase approximation~\cite{Aryasetiawan_PRB2004,Imada_JPSJ2010} implemented in RESPACK~\cite{K.Nakamura2021}. 
We set the cutoff energy of the polarization function to $5.0$ Ry. 
Details of the transfer integrals and interaction parameters 
are summarized in the Supplemental Material~\cite{supplementary}. 
{We note that the difference in the intra-dimer hoppings $t_{1}$ and
$t_2$ and the existence of the site potential $\delta$ indicate the inequivalence of the dimers.}
In actual calculations, we subtract a constant value $\Delta_{\rm DDF}$
from onsite and offsite Coulomb interactions to consider  
the interlayer screening ~\cite{Nakamura_JPSJ2010,K.Nakamura2012}.
Following previous studies~\cite{K.Nakamura2012,Ido_npjQ2022,D.Ohki2023,Yoshimi_PRL2023}, 
we employ $\Delta_{\rm DDF}=0.2$ eV.
We confirm that the results are insensitive to $\Delta_{\rm DDF}$. 
We also perform an electron-hole transformation to reduce the numerical cost. 

{\it many-variable variational Monte Carlo (mVMC) analysis.---}
The effective Hamiltonian is solved 
using the mVMC method~\cite{Tahara_2008JPSJ1,T.Misawa2019},
which can take into account quantum fluctuations and spatial correlations seriously.
The trial wave function used in this study is given by 
\begin{align}
&\ket{\psi}=\mathcal{P}_{G}\mathcal{P}_{J}\mathcal{L}_{S}\ket{\phi_{\rm pair}}, \\
&\mathcal{P}_{G}={\rm exp}\left[\sum_{i}g_{i}n_{i\uparrow}n_{i\downarrow}\right], \\
&\mathcal{P}_{J}={\rm exp}\left[\frac{1}{2}\sum_{i\neq j}v_{ij}N_{i}N_{j}\right], \\
&\ket{\phi_{\rm pair}}=\left[\sum^{N_{\rm site}}_{i,j}f_{ij}c^{\dagger}_{i\uparrow}c^{\dagger}_{j\downarrow}\right]^{N_{\rm e}/2}\ket{0},
\end{align}
where $\mathcal{P}_{G}$, $\mathcal{P}_{J}$, and $\mathcal{L}_{S}$ are the Gutzwiller factor~\cite{Gutzwiller_PRL1963}, long-range Jastrow factor~\cite{Jastrow_PR1955,Capello_PRL2005}, and total spin projector~\cite{RingShuck,Mizusaki_PRB2004}, respectively~\cite{Becca_TB2017,T.Misawa2019}. 
$N_{\rm e}$ and $N_{\rm site}$ indicate the number of electrons and sites, respectively. 
We impose a $1\times 4$ sublattice structure on variational parameters.
We use the Hartree-Fock approximation results as the initial $f_{ij}$ values. 
The spin-singlet projection ($S=0$) is used in ground-state calculations for ${L_a=L_b}\leq 8$, while it is not used for ${L_a=L_b}\geq 10$
to reduce the numerical costs. We confirm that the spin projection does not largely affect physical quantities, such as the spin structure factors.
All variational parameters are simultaneously optimized 
using the stochastic reconfiguration method~\cite{Sorella_PRB2001}.

\begin{figure}[t]
\begin{center}
\includegraphics[width=85mm]{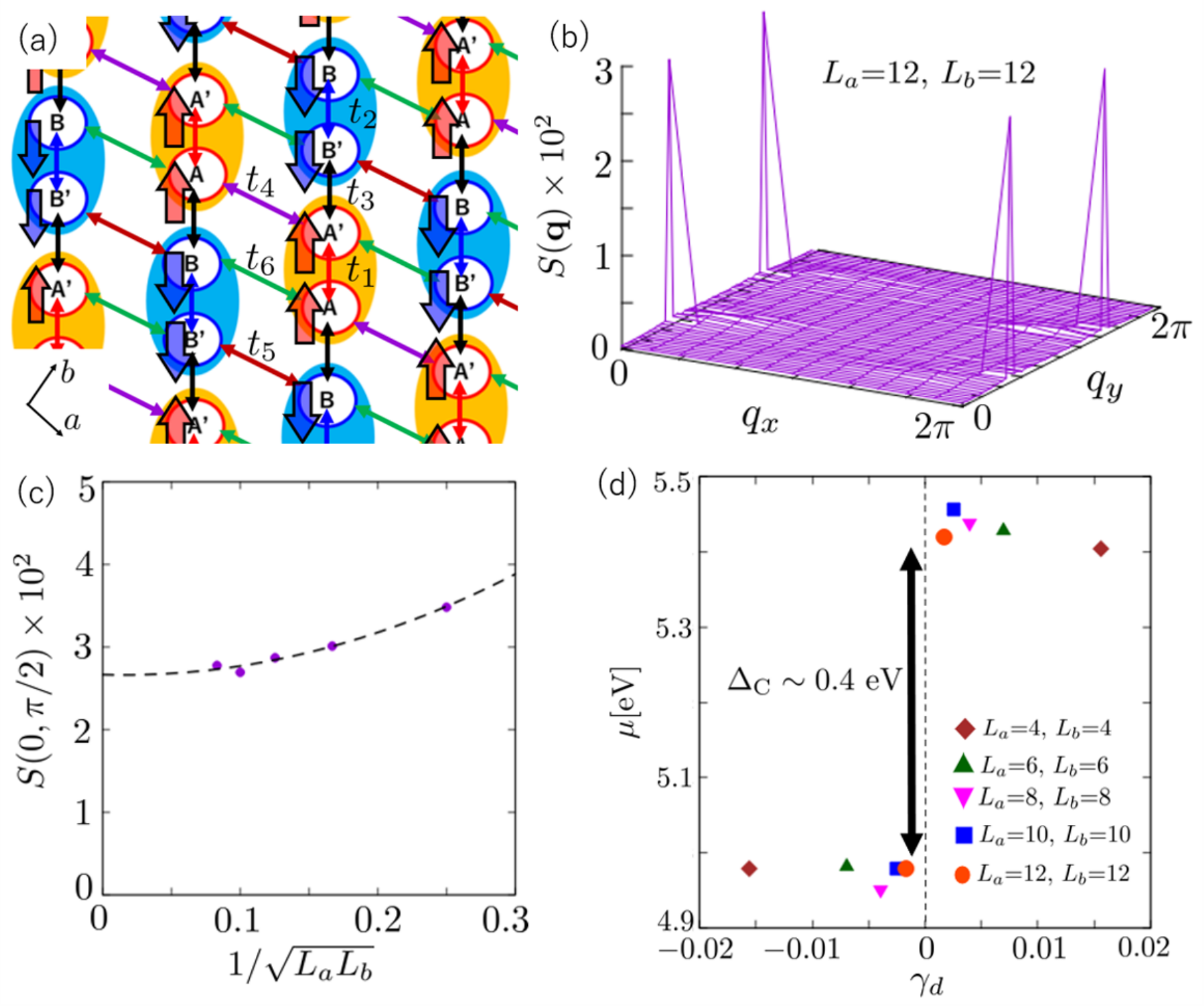}
\end{center}
\caption{(a)~Schematic of the DAF state and representative hopping integrals ($t_1$-$t_6$)
in the conducting layer of (EDO-TTF-I)$_{2}$ClO$_{4}$.
(b)~Spin structure factor obtained by the mVMC method for ${L_a=L_b}=12$. Sharp peaks
appear at $\vec{q}=(0,\pi/2), (0,3\pi/2)$, which correspond to the DAF state.
(c)~Size dependence of the peak value of the spin structure factors.
(d)~Doping dependence of the chemical potential. 
}
\label{Fig3}
\end{figure}

As detailed in the Supplementary Materials, 
the Hartree-Fock approximation shows that the DAF state (Fig.\ref{Fig3} (a)) 
and antiferromagnetic states with charge ordering (AF+CO) are
ground state candidates.
Within the Hartree-Fock approximation, the ground state of 
the effective Hamiltonian is the AF+CO state. 
However, using the mVMC method, we find that the 
DAF state becomes the ground state of the effective Hamiltonian.
We also find that the AF+CO state converges to the DAF state
after optimization, even when using the variational parameters of AF+CO state as the initial state.
This result indicates that correlation effects beyond the mean-field approximation are important in stabilizing the DAF state.

To examine the existence of long-range antiferromagnetic order,
we calculate the spin-correlation functions of the ground state, defined as
\begin{eqnarray}
\label{SqNq}
S(\vec{q})=\frac{1}{(N_{\rm site})^{2}}\sum_{i,j}\left<\vec{S}_{i}\cdot \vec{S}_{j}\right>e^{i\vec{q}\cdot (\vec{r}_{i}-\vec{r}_{j})}, 
\end{eqnarray}
where the original lattice structure is mapped to the equivalent ${L_a \times 4L_b}$ 
square lattice for simplicity.    
Figure \ref{Fig3} (b) shows $S(\vec{q})$ in the momentum space, 
with sharp Bragg peaks at $(q_x,q_y)$=$(0,\pi/2)$ and $(0,3\pi/2)$. 
As shown in Fig.~\ref{Fig3} (c), we confirm that the peak values of $S(\vec{q})$ remain finite in the thermodynamic limit. 
We also confirm that charge densities are uniform and there is no signature of a charge-ordered state.
These results show that the ground state of the effective Hamiltonian is the DAF state.
In addition, we calculate the charge gap $\Delta_{\rm c}$,
given by $\Delta_{\rm c}$ = $\mu(N_{\rm e}+1)-\mu(N_{\rm e}-1)$, 
where the chemical potential is defined as $\mu(N_{\rm e}+1)$ = $[E(N_{\rm e}+2)-E(N_{\rm e})]/2$. 
Figure \ref{Fig3} (d) shows the doping rate $\gamma_{d}$ = $N_{\rm e}/N_{\rm site}-1.5$ dependence of
the chemical potential. 
From this plot, we estimate the charge gap [$\Delta_{\rm c}$ = $\mu(N_{\rm e}+1)-\mu(N_{\rm e}-1)$] as
$\Delta_{\rm c}\sim 0.4$ eV. This result indicates that the ground state of 
(EDO-TTF-I)$_2$ClO$_4$ is the DAF insulator.

{\it Spin splitting.---}Based on the results obtained using the mVMC method, 
we analyze spin splitting in the DAF state using the Hartree-Fock approximation
 (for more details, see Ref.~\cite{supplementary}).
We assume the DAF order and scale the interaction parameters
to reproduce the charge gap $\Delta_{\rm c}\sim0.4$ eV 
obtained from the mVMC calculations. The scaling ratio $\lambda$,
which monotonically scales the onsite and offsite Coulomb interactions,
is estimated to be $\lambda=0.7$. The details of the Hartree-Fock calculations are provided in the Supplemental Material.

\begin{figure}[tb]
\begin{center}
\includegraphics[width=85mm]{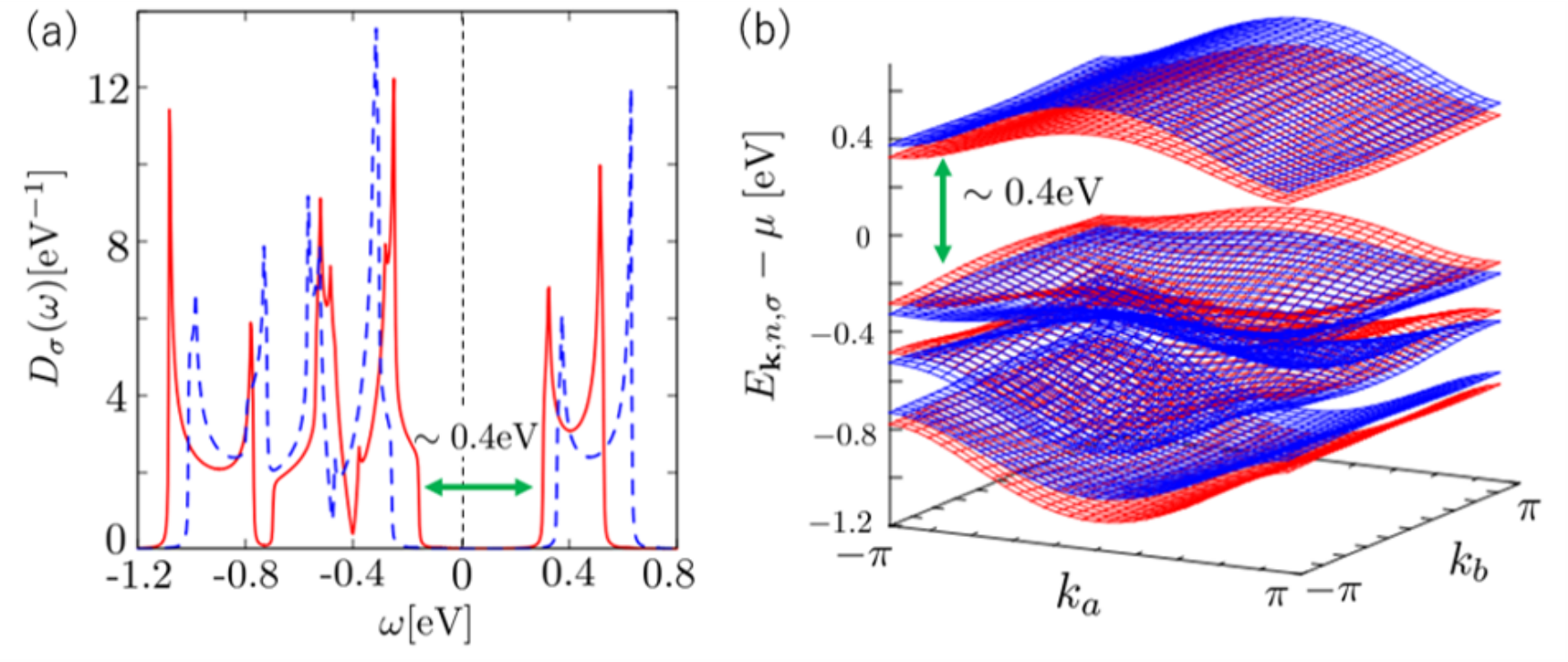}
\end{center}
\caption{(a) Density of state (DOS) of the low-energy effective Hamiltonian (EDO-TTF-I)$_2$ClO$_4$ for the DAF state obtained by the Hartree-Fock approximation. DOS for up (down) spin is described by
the red (blue) lines.
(b) Band dispersions of the DAF state obtained by the Hartree-Fock approximation.  
The red (blue) surfaces describe up-spin (down-spin) band dispersions.}
\label{Fig4}
\end{figure}

Using the Hartree-Fock approximation, we calculate the {DOS}
defined as $D_{\sigma}(\omega)$=$(\pi {L_aL_b} )^{-1}\sum_{\vec{k},n}{\rm Im}(\omega-i\eta-E_{\vec{k},n,\sigma}+\mu)^{-1}$, 
where $\mu$ and $\eta$ are the chemical potential and the smearing factor, respectively.
$E_{\vec{k},n,\sigma}$ denotes the $n$-th eigenvalue of the mean-field 
Hamiltonian at momentum $\vec{k}$. 
We set $\eta=0.002$~eV.
{As shown in Fig.~\ref{Fig4}~(a),
spin splitting 
{occurs in the}
DAF order ($D_{\uparrow}(\omega)\ne D_{\downarrow}(\omega)$).} 
The electronic band dispersions in Fig.~\ref{Fig4}(b)
also show isotropic spin splitting over the entire Brillouin zone.
This demonstrates that (EDO-TTF-I)$_{2}$ClO$_{4}$ 
can be fully compensated ferrimagnets if a DAF order occurs.

Here, we analyze the origin of spin splitting in (EDO-TTF-I)$_{2}$ClO$_{4}$.
As in the case of the simple model, 
$t_{-}=(t_{1}-t_{2})/2$ and $\delta$ can induce spin splitting.
From the $ab$ $initio$ calculations, we find that
$t_{-}=0.036$ eV is comparable to $\delta=0.047$ eV. 
Thus, spin splitting in (EDO-TTF-I)$_{2}$ClO$_{4}$ is induced by both $t_{-}$ and $\delta$.
One might think that finite $\delta$ would make the total magnetization finite; 
however, a charge gap guarantees that the total magnetization is robust against perturbations. 
In this case, the total magnetization is zero at $\delta=0$ 
and remains zero even if $\delta$ is added,
provided that $\delta$ is significantly smaller than the charge gap. 

{\it Summary and discussion.---} In this Letter, we present a simple method to realize fully compensated
ferrimagnetism using the dimer degrees of freedom, which are typical of organic compounds.
Using a simple model, we demonstrate that the inequivalence of the two dimers and DAF order {can induce} fully compensated ferrimagnets at commensurate filling. Furthermore, $ab$ $initio$ calculations suggest that
the ground state of (EDO-TTF-I)$_2$ClO$_4$
is a DAF insulator with inequivalent dimers. 
As a result, the DAF order induces isotropic spin splitting
in the electronic band structure. 
Our study shows that the key to realizing compensated ferrimagnetism lies in inequivalent dimer structures induced by anion ordering. 
This finding offers an unanticipated direction in materials design, 
where exotic magnetism can be achieved by selecting and modifying anions to exhibit anion ordering.
The discovery of compensated ferrimagnetism in inequivalent dimer structures, 
as well as the potential for materials design using anion ordering, 
demonstrates that organic compounds offer a versatile platform for realizing exotic magnetism.
An intriguing future issue would be to examine the doping effects in the DAF insulating state. 
Because the lowest unoccupied band is fully polarized and
its DOS is large (Fig.\ref{Fig4}~(a)), unconventional superconductivity, such as triplet superconductivity, is expected ~\cite{Mazin_arXiv2022}.
Further experimental and theoretical investigations in this direction are 
desirable.

\begin{acknowledgements}  
The authors thank Y. Nakano, M. Ishikawa, H. Yamochi, and A. Otsuka for the fruitful discussions.
This work was financially supported by Grants-in-Aid for Scientific Research (KAKENHI) 
(Grant Nos. 23H03818, 23KJ1065, 15K05166, 22K18683 21H01793), 
Grant-in-Aid for Scientific Research for Transformative Research Areas (A) ``Condensed Conjugation'' (No. JP20H05869) from Japan Society for the Promotion of Science (JSPS), and JST SPRING (Grant No. JPMJSP2125). The computations were performed using the facilities at the Supercomputer Center, Institute for Solid State Physics, University of Tokyo.
\end{acknowledgements}

\clearpage
\noindent

{\Large
Supplemental Material for ``Compensated Ferrimagnets with Colossal Spin Splitting in Organic Compounds"
}

\section{Parameters in ab initio low-energy effective Hamiltonians}
\begin{figure}[htb]
\begin{center}
\includegraphics[width=80mm]{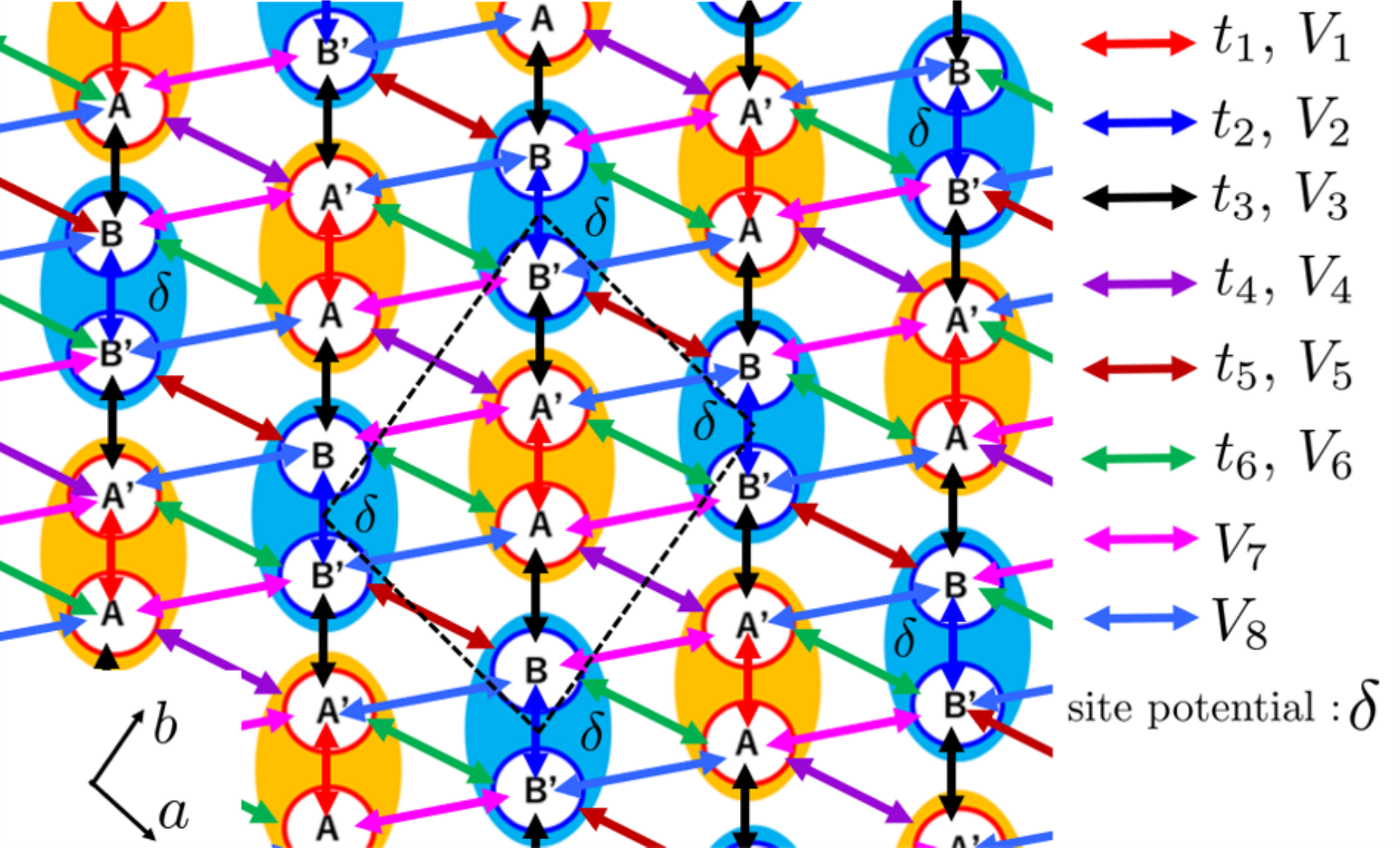}
\end{center}
\caption{Schematic of the transfer integrals 
and the off-site Coulomb interactions in the conduction layer. 
}
\label{nw}
\end{figure}

We calculate the electronic band structures by the density functional theory (DFT) 
using Quantum ESPRESSO~\cite{P.Giannozzi2017} and evaluate the transfer integrals by the maximally localized Wannier functions (MLWFs). Then, we evaluate Coulomb interaction by the constraint random phase approximation (cRPA) using RESPACK~\cite{K.Nakamura2021}. 
The values of the transfer integrals larger than $0.020$~eV and the Coulomb interactions are shown in Table~\ref{combinedTable}. We show schematic illustrations of the transfer integrals and the
Coulomb interactions in Fig.~\ref{nw}.
All input and output files are uploaded to the ISSP data repository~\cite{datarepo}.

\begin{table}[hb]
\begin{tabular}{clcl} 
\hline
\hline
Transfer integrals [eV] & &~~~~~~ Coulomb interactions [eV] & \\
\hline
$\delta$ & 0.047 &~~~~~~ $U_{A}$=$U_{A^{\prime}}$ & 2.094\\

        &        & ~~~~~~ $U_{B}$=$U_{B^{\prime}}$ & 2.076\\
$t_{1}$ & 0.252  &~~~~~~ $V_{1}$ & 0.903\\
$t_{2}$ & 0.179  &~~~~~~ $V_{2}$ & 0.884\\
$t_{3}$ & 0.128  &~~~~~~ $V_{3}$ & 0.880\\
$t_{4}$ & 0.112  &~~~~~~ $V_{4}$ & 0.700\\
$t_{5}$ & 0.084  &~~~~~~ $V_{5}$ & 0.681\\
$t_{6}$ & 0.058  &~~~~~~ $V_{6}$ & 0.614\\
$t_{7}$ & -      &~~~~~~ $V_{7}$ & 0.659\\
$t_{8}$ & -      &~~~~~~ $V_{8}$ & 0.628\\
\hline
\hline
\end{tabular}
\caption{Transfer integrals and screened Coulomb interactions of the $ab$ $initio$ low-energy effective Hamiltonians for (EDO-TTF-I)$_{2}$ClO$_{4}$.}
\label{combinedTable}
\end{table}

\clearpage
\section{Partial densities of states}
\begin{figure}[t]
\begin{center}
\includegraphics[width=80mm]{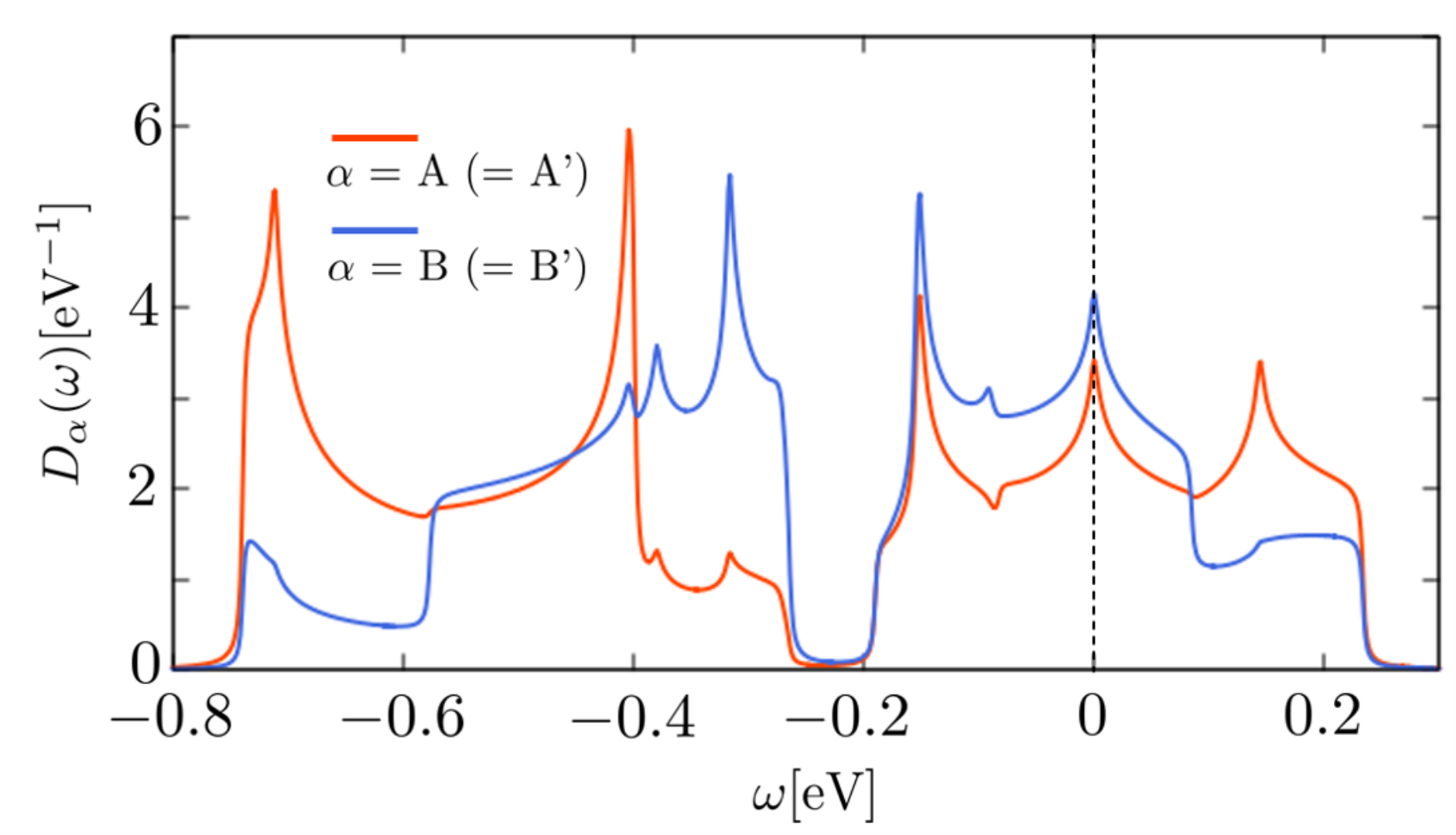}
\end{center}
\caption{{PDOS obtained by the tight-binding model. Inequivalence of the dimers appears in the PDOS ($D_{A}(\omega) \ne D_{B}(\omega)$). }
}
\label{PDOS}
\end{figure}

{
{To see the inequivalence of the dimers,} we calculate the partial densities of {states}~(PDOS) using the {one-body part of the low-energy effective Hamiltonians for (EDO-TTF-I)$_{2}$ClO$_{4}$}
in {the} momentum space, which is given by  
\begin{equation}
\begin{split}
\label{tight-binding}
{H^{0}_{\sigma}(\vec{k})}&=\sum_{\vec{k}}\sum_{\Delta\vec{r},\alpha,\beta}t_{\Delta\vec{r},\alpha\beta}e^{i\vec{k}\cdot\Delta\vec{r}}c^{\dagger}_{\vec{k},\alpha,\sigma}c_{\vec{k},\beta,\sigma} \\
&=\sum_{\vec{k}}\sum_{\Delta\vec{r},\alpha,\beta}H^{0}_{\alpha\beta,\sigma}(\Vec{k})c^{\dagger}_{\vec{k},\alpha,\sigma}c_{\vec{k},\beta,\sigma}.
\end{split}  
\end{equation}
Here, $t_{\Delta\vec{r},\alpha\beta}$ is the transfer integrals obtained by RESPACK
{and} $\Delta\vec{r}$ 
{represents} the translational vector. 
{The Hamiltonian} ${H}^{0}_{\sigma}(\Vec{k})$ satisfies the following eigenvalue {equation:}
\begin{equation}
\label{eigen_Eq}
{H}^{0}_{\sigma}(\vec{k})\ket{{\vec{k},n,\sigma}}=E_{n,\sigma}(\vec{k})\ket{\vec{k},n,\sigma},
\end{equation}
\begin{equation}
\label{eigen_vec}
\ket{{\vec{k},n,\sigma}}=
\left(
\begin{array}{c}
d_{A,n,\sigma}(\vec{k}) \\
d_{A^{\prime},n,\sigma}(\vec{k}) \\
d_{B,n,\sigma}(\vec{k}) \\
d_{B^{\prime},n,\sigma}(\vec{k})  \\
\end{array}
\right),
\end{equation}
{where} $E_{n,\sigma}(\vec{k})$ and $\ket{\Vec{k},n,\sigma}$ are the eigenvalue and {eigenvectors} of ${H}^{0}$. 
{The band and spin indices are represented by $n$ and $\sigma$, respectively.}
Using $E_{n,\sigma}(\vec{k})$ and $d_{\alpha,n,\sigma}(\textbf{k})$, the PDOS $D_{\alpha}(\omega)$ {can be calculated as} 
\begin{equation}
\label{pdos_eq}
D_{\alpha}(\omega)=\frac{1}{\pi {L_a L_b}}\sum_{\vec{k},n,\sigma}{\rm Im}\frac{|d_{\alpha,n,\sigma}(\vec{k})|^{2}}{\omega-i\eta-E_{\vec{k},n,\sigma}+\mu},
\end{equation}
where $\eta$ takes the positive infinitesimal value and $\mu$ is the chemical potential determined  
{to set the electron number to 6.}
We set $\eta$ = $0.002$ eV in the numerical calculation. Figure \ref{PDOS} shows {the PDOS} $D_{\alpha}(\omega)$.
{We consider the paramagnetic state in this calculation.}
{The inequivalence in $D_{A}(\omega)$ and $D_{B}(\omega)$
indicates the inequivalence 
of the dimer I and II.}
{We note that the relations} $D_{A}(\omega)$ = $D_{A'}(\omega)$ and $D_{B}(\omega)$ = $D_{B'}(\omega)$ are satisfied due to space-inversion symmetry.}

\clearpage
\section{Details of the Hartree-Fock approximation}
\begin{figure*}[htb]
\begin{center}
\includegraphics[width=140mm]{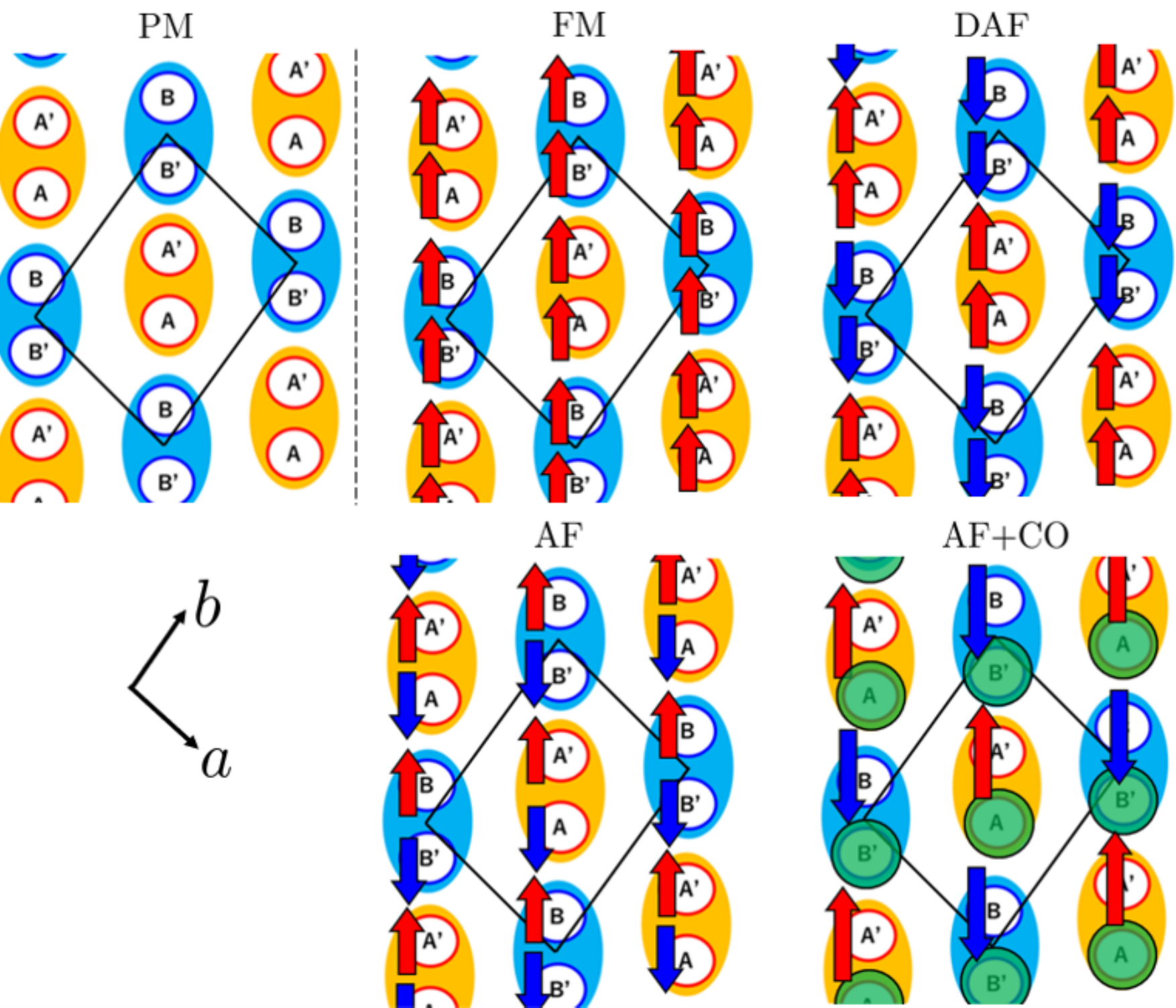}
\end{center}
\caption{Schematic illustrations of initial states used in the HF approximation. The up and down arrows indicate the spin-up and spin-down states, respectively. The orange and blue ellipses represent the dimer states constructed by A-A$^{\prime}$ molecules and B-B$^{\prime}$ molecules, respectively. The green circles represent the charge-rich sites.}
\label{initial}
\end{figure*}

To examine the ground state candidates of (EDO-TTF-I)$_2$ClO$_4$,
we perform the Hartree-Fock (HF) approximation to  
the $ab$ $initio$ low-energy effective Hamiltonians defined in Eq.~(5) in the main text.
We use the unrestricted HF code implemented in mVMC~\cite{T.Misawa2019}. Using the results of HF calculations, we generate the initial variational parameters for the mVMC method.
In this study, we examine the ordered states with $\vec{q}=0$, i.e., we consider the symmetry-broken states within the unit cell.
As illustrated in Fig. \ref{initial}, we consider five initial states: (i) the paramagnetic (PM) state, (ii) the ferromagnetic (FM) state, (iii) the dimer antiferromagnetic (DAF) state, (iv) the AF state, (v) the AF state with charge ordering (AF+CO).  
To examine the correlation effects, we introduce the parameter $\lambda$, which monotonically scales the Coulomb interactions
as {$\tilde{U}_{i\alpha}\equiv \lambda(U_{i\alpha}-\Delta_{\rm DDF})$} and {$\tilde{V}_{i\alpha j\beta}\equiv \lambda(V_{i\alpha j\beta}-\Delta_{\rm DDF})$}.
Here, $\Delta_{\rm DDF}$ denotes the constant shift that takes into account the  
screening effects between conduction layers~\cite{Nakamura_JPSJ2010,K.Nakamura2012}.

\begin{figure*}[htb]
\begin{center}
\includegraphics[width=120mm]{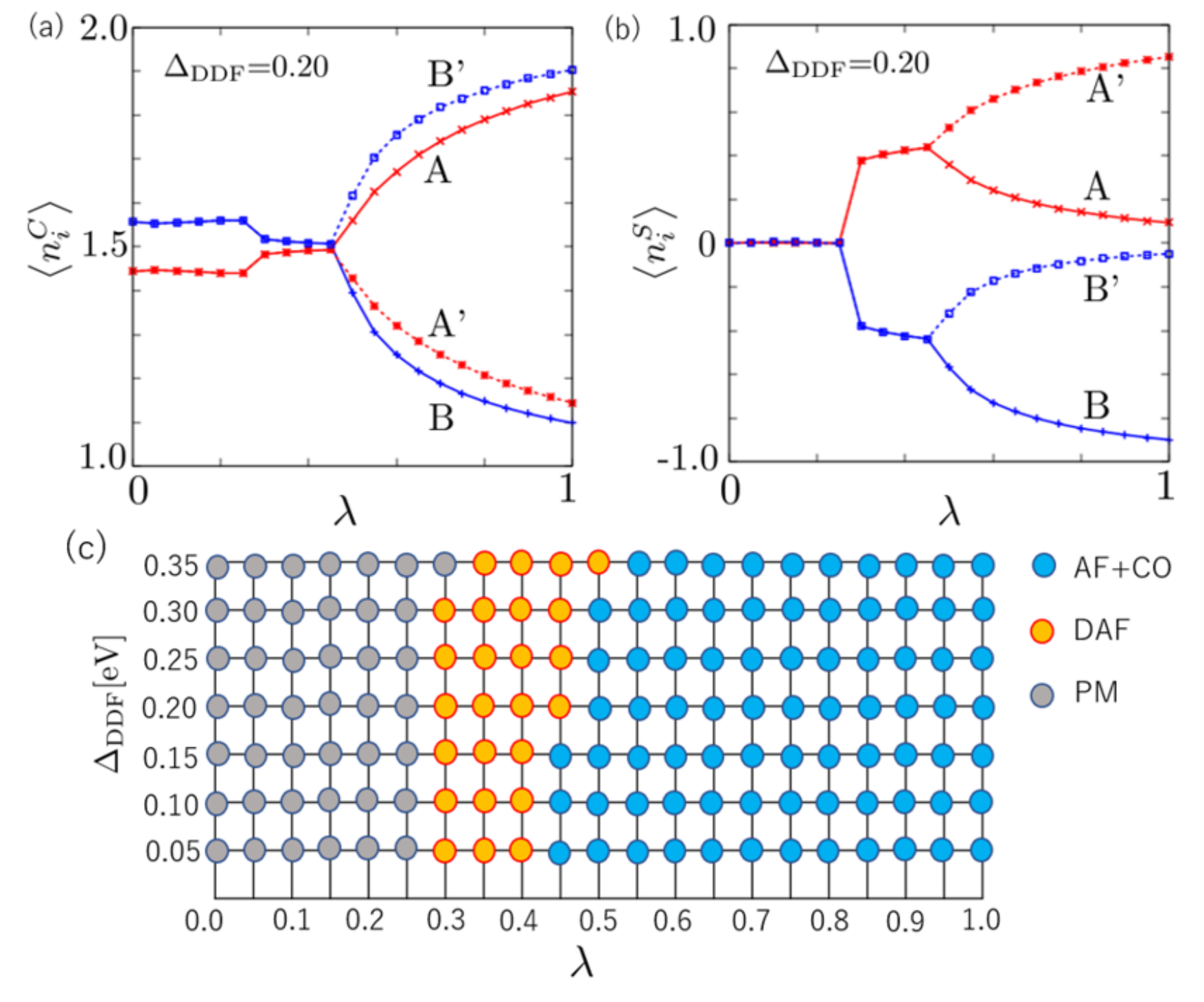}
\end{center}
\caption{$\lambda$ dependence of (a) the charge density and (b) the spin density. (c) $\lambda$-$\Delta_{\rm DDF}$ phase diagram obtained by the HF approximation.}
\label{HF}
\end{figure*}

Figures \ref{HF} (a) and (b) show the charge density $\left<n^{\rm C}_{i}\right>=\left<n_{i\uparrow}+n_{i\downarrow}\right>$ and the spin density $\left<n^{\rm S}_{i}\right>=\left<n_{i\uparrow}-n_{i\downarrow}\right>$ at the $i$th (=A, A$^{\prime}$, B, B$^{\prime}$) site in the ground states obtained by the HF approximation for $\Delta_{\rm DDF}=0.2$~eV, respectively. We find that the PM state [$\langle n^{\rm C}_{A(B)}\rangle=\langle n^{\rm C}_{A'(B')} \rangle$ and $\langle n^{\rm S}_{i}\rangle=0$] is the ground state below $\lambda\sim 0.3$. 
In $0.3\lesssim\lambda\lesssim0.5$, DAF state [$\langle n^{\rm C}_{A(B)}\rangle=\langle n^{\rm C}_{A'(B')} \rangle$ and $\langle n^{\rm S}_{A}\rangle=\langle n^{\rm S}_{A'}\rangle=-\langle n^{\rm S}_{B}\rangle=-\langle n^{\rm S}_{B'}\rangle$] becomes the ground state. For $\lambda \gtrsim 0.5$, the AF state with the CO  [$\langle n^{\rm C}_{A(B')}\rangle > \langle n^{\rm C}_{A'(B)} \rangle$ and $\langle n^{\rm S}_{A'(B')}\rangle > \langle n^{\rm S}_{A(B)}\rangle$] becomes the ground state.
Figure \ref{HF} (c) shows the $\lambda$-$\Delta_{\rm DDF}$ phase diagram.
Since amplitudes of the on-site and off-site Coulomb interactions become small
with increasing $\Delta_{\rm DDF}$, the  
phase boundary between the DAF state and the AF+CO state 
slightly shifts to large $\lambda$ region by increasing $\Delta_{\rm DDF}$.

\clearpage
\section{mVMC analysis of the ground-state phase diagram}
\begin{figure*}[t]
\begin{center}
\includegraphics[width=120mm]{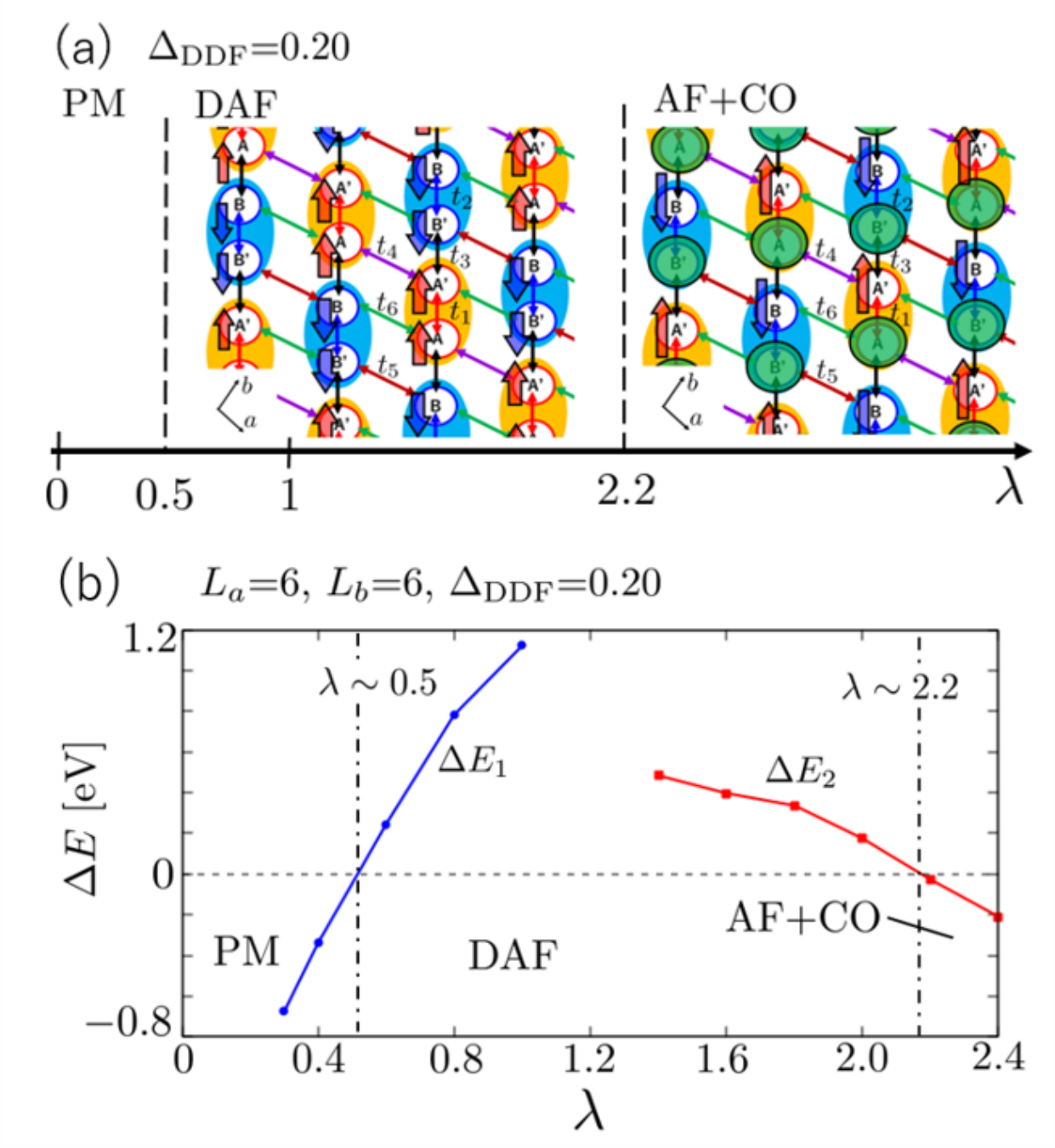}
\end{center}
\caption{(a)~The ground state phase diagram as a function of $\lambda$ obtained by the mVMC calculation.
(b)~$\lambda$-dependencies of the energy difference between the PM state and the DAF state ($\Delta E_{1}$) and the one between the AF+CO state and the DAF state ($\Delta E_{2}$).} 
\label{pd_mVMC}
\end{figure*}

\begin{figure*}[t]
\begin{center}
\includegraphics[width=140mm]{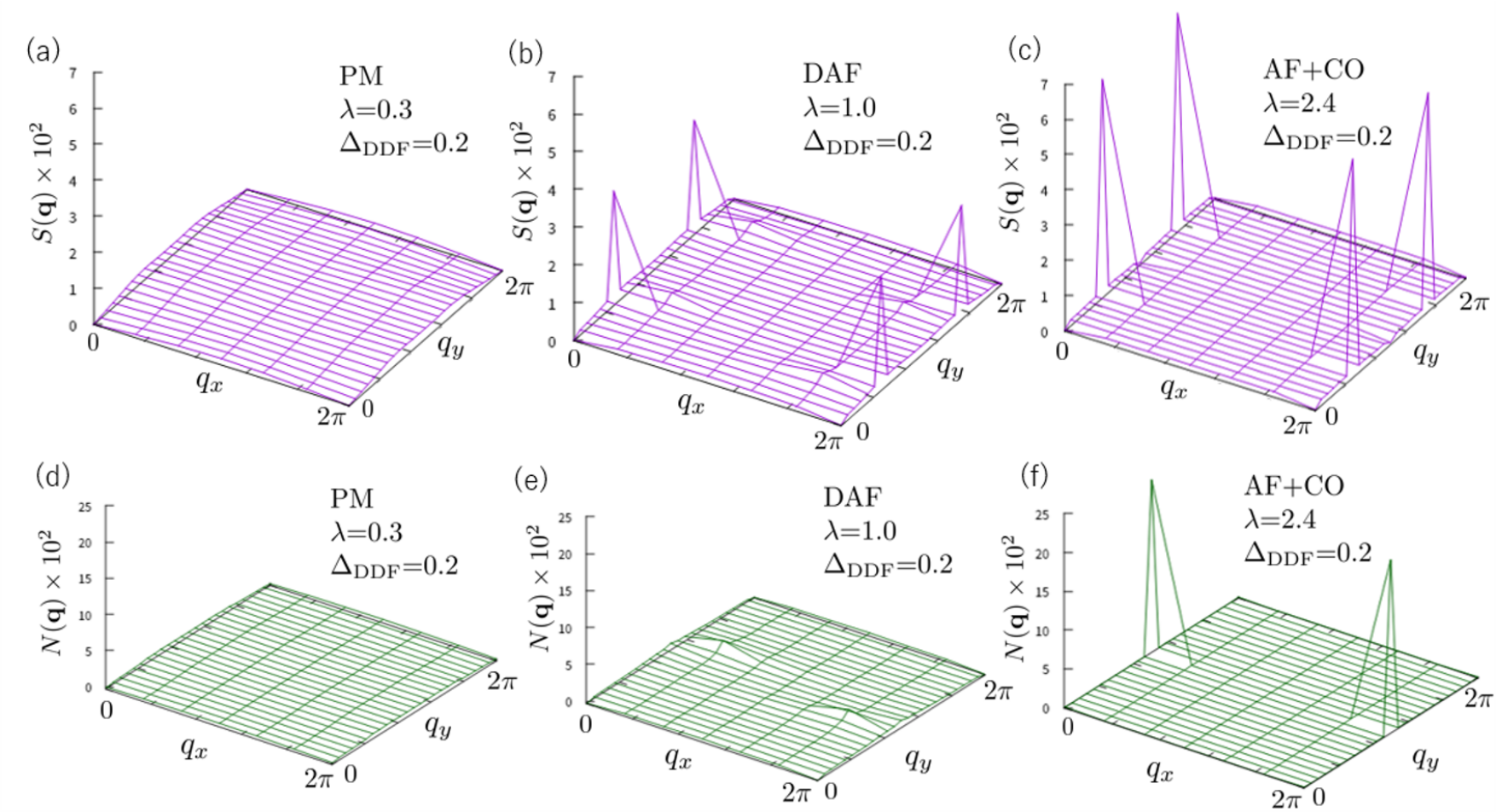}
\end{center}
\caption{Spin structure factors $S(\vec{q})$ [charge structure factors $N(\vec{q})$ ] of the PM state ($\lambda$ = $0.3$), the DAF state ($\lambda$ = $1$), and the  AF+CO state ($\lambda$ = $2.4$) in (a), (b), and (c) [(d), (e), and (f)], respectively.}
\label{sqnq}
\end{figure*}

Based on the results obtained by the HF calculations, we investigate the ground states using the mVMC method, which can treat correlation effects more accurately. 
Following previous studies~\cite{K.Nakamura2012,Ido_npjQ2022,D.Ohki2023,Yoshimi_PRL2023}, we 
take $\Delta_{\rm DDF}=0.20$~eV.  
Figure~\ref{pd_mVMC}(a) shows the phase diagram as a function of $\lambda$.
By increasing $\lambda$, the phase transition between the DAF state and the PM state occurs around $\lambda=0.5$. 
Above $\lambda\sim 2.2$, the AF+CO state becomes the ground state due to the off-site Coulomb interactions.
Figure~\ref{pd_mVMC}(b) shows the energy difference between the PM state (the AF+CO state)  and the DAF state, i.e., $\Delta{E}_{1}=E_{\rm PM}-E_{\rm DAF}$
($\Delta E_{2}=E_{\rm AF+CO}-E_{\rm DAF}$) as a function of $\lambda$ 
for ${L_a=L_b}=6$ lattice.  
The AF+CO state is a quasi-stable state for $\lambda\gtrsim 1.4$ and
becomes the ground state for $\lambda\gtrsim 2.2$.
We note that the AF+CO state is not stabilized even when we select the AF+CO state as an initial state for $\lambda\lesssim 1.2$.
Figure \ref{sqnq} shows the spin and {charge} density structure factors of the PM, the DAF, and the AF+CO states. 
The charge structure factor is defined by
\begin{align}
\label{SqNq}
&N(\vec{q})=\frac{1}{(N_{\rm site})^{2}}\sum_{i,j}\left<(N_{i}-\bar{N})\cdot (N_{j}-\bar{N})\right>e^{i\vec{q}\cdot (\vec{r}_{i}-\vec{r}_{j})}, \\
&\bar{N}=\frac{1}{N_{\rm site}}\sum_{i}\left<N_{i}\right>.
\end{align}
The ordering wave vectors $\vec{q}=(0,\pi/2), (0,3\pi/2)$ in $S(\vec{q})$ (Fig. \ref{sqnq} (b)) correspond to the DAF state. 
Meanwhile, the ordering wave vectors $\vec{q}=(0,\pi/2), (0,3\pi/2)$ in $S(\vec{q})$ and $\vec{q}=(0,\pi)$ in $N(\vec{q})$ (Fig.~\ref{sqnq}(c) and (f)) correspond to the AF+CO state.

\clearpage
\section{Mean-filed Hamiltonian in the momentum space}
To see the spin splitting of the DAF states, we calculate band dispersions and 
{DOS} using the one-body Green functions  
obtained by the HF approximation. Here, we set $\lambda=0.7$, which reproduces the charge gap estimated by the mVMC method. 
By performing the Fourier transformation for
the mean-field Hamiltonian in the real space,
we obtain the mean-field Hamiltonian in the momentum space, which is given by
 \begin{equation}
 \begin{split}
 \label{HF_k}
 H&=\sum_{\vec{k}}\sum_{\Delta\vec{r},\alpha,\beta,\sigma}t_{\Delta\vec{r},\alpha\beta}e^{i\vec{k}\cdot\Delta\vec{r}}c^{\dagger}_{\vec{k},\alpha,\sigma}c_{\vec{k},\beta,\sigma} \\
 &+\sum_{\vec{k}}\sum_{\alpha,\sigma}U_{\alpha}\left<n_{\alpha,\bar{\sigma}}\right>c^{\dagger}_{\vec{k},\alpha,\sigma}c_{\vec{k},\alpha,\sigma} \\
 &+\sum_{\vec{k}}\sum_{\Delta\vec{r},\alpha,\beta,\sigma}V_{\Delta\vec{r},\alpha\beta}\left<N_{\beta}\right>c^{\dagger}_{\vec{k},\alpha,\sigma}c_{\vec{k},\alpha,\sigma} \\ 
 &-\sum_{\vec{k}}\sum_{\Delta\vec{r},\alpha,\beta,\sigma}V_{\Delta\vec{r},\alpha\beta}\left<c^{\dagger}_{\vec{r}_{0}+\Delta\vec{r},\beta,\sigma} c_{\vec{r}_{0},\alpha,\sigma}\right>e^{i\vec{k}\cdot\Delta\vec{r}}c^{\dagger}_{\vec{k},\alpha,\sigma}c_{\vec{k},\beta,\sigma}.
 \end{split}
 \end{equation}
 Here, $\Delta\vec{r}$ denotes the translational vector and $\bar{\sigma}=-\sigma$. 
 Off-diagonal one-body Green functions in the real space
 are represented by $\langle c^{\dagger}_{\vec{r}_{0}+\Delta\vec{r},\beta,\sigma} c_{\vec{r}_{0},\alpha,\sigma}\rangle$.  
 We take $\vec{r}_{0}$ = $\vec{0}$ as the representative coordinate of $\vec{r}_{0}$ because $\vec{r}_{0}$ is an arbitrary coordinate due to translational symmetry. Using the mean-field Hamiltonians in the momentum space,
we calculate the band dispersions and the {DOS}. 
All values of the one-body Green functions are  
uploaded to the ISSP data repository~\cite{datarepo}.

\providecommand{\noopsort}[1]{}\providecommand{\singleletter}[1]{#1}%

\end{document}